\documentclass[conference]{IEEEtran}
\usepackage{ifpdf}
\usepackage{cite}
\usepackage{graphicx}
\usepackage[cmex10]{amsmath}
\usepackage{amssymb}
\usepackage{array}
\usepackage{bm}
\usepackage{mathtools}
\usepackage[caption=false,font=footnotesize]{subfig}
\usepackage{booktabs}
\usepackage{color}

\begin{document}
\title{
    Vision-Aided Frame-Capture-Based\\ 
    CSI Recomposition for WiFi Sensing:\\
    A Multimodal Approach
}
\author{
    \IEEEauthorblockN{
        \normalsize Hiroki Shimomura\IEEEauthorrefmark{1},
        \normalsize Yusuke Koda\IEEEauthorrefmark{2}\IEEEauthorrefmark{5},
        \normalsize Takamochi Kanda\IEEEauthorrefmark{1},
        \normalsize Koji Yamamoto\IEEEauthorrefmark{1}\IEEEauthorrefmark{6}, 
        \normalsize Takayuki Nishio\IEEEauthorrefmark{1}\IEEEauthorrefmark{3}, and
        \normalsize Akihito Taya\IEEEauthorrefmark{4}
    }
    \IEEEauthorblockA{
        \IEEEauthorrefmark{1}\small Graduate School of Informatics, Kyoto University, Yoshida-honmachi, Sakyo-ku, Kyoto, 606-8501, Japan \\
        \IEEEauthorrefmark{2}\small Centre for Wireless Communications, University of Oulu, 90014 Oulu, Finland \\
        \IEEEauthorrefmark{3}\small School of Engineering, Tokyo Institute of Technology, Ookayama, Meguro-ku, Tokyo, 152-8550, Japan \\
        \IEEEauthorrefmark{4}\small Department of Integrated Information Technology, Aoyama Gakuin University, \\Fuchinobe, Chuo-ku, Sagamihara-shi, Kanagawa, 252-5258, Japan\\
        \IEEEauthorrefmark{5}\small koda@ieee.org,
        \IEEEauthorrefmark{6}\small kyamamot@i.kyoto-u.ac.jp
    }
}

\maketitle

\begin{abstract}
Recompositing channel state information (CSI) from the beamforming feedback matrix (BFM), which is a compressed version of CSI and can be captured because of its lack of encryption, is an alternative way of implementing firmware-agnostic WiFi sensing.
In this study, we propose the use of camera images toward the accuracy enhancement of CSI recomposition from BFM.
The key motivation for this vision-aided CSI recomposition is to draw a first-hand insight that the BFM does not fully involve spatial information to recomposite CSI and that this could be compensated by camera images.
To leverage the camera images, we use multimodal deep learning, where the two modalities, i.e., images and BFMs, are integrated to recomposite the CSI.
We conducted experiments using IEEE 802.11ac devices.
The experimental results confirmed that the recomposition accuracy of the proposed multimodal framework is improved compared to the single-modal framework only using images or BFMs.
\end{abstract}

\section{Introduction}
\label{sec:intro}
Currently, channel state information (CSI) in wireless local area networks (WLANs) has attracted increasing attention because of its fine-grained propagation characteristics.
While CSI-based sensing has been widely studied, obtaining CSI from off-the-shelf WLAN devices requires customized firmware, whose custom can be applied only to a specific chipset and protocol \cite{halperin2011tool,gringoli2019free}.
This is because off-the-shelf devices discard CSI after the beamforming operation in physical layer components.
This constrains the range of CSI-based sensing applications.

To alleviate this limitation, some studies \cite{miyazaki2019initial,kanda2022respiratory} utilized the beamforming feedback matrix (BFM), which is a series of right singular matrices for CSI matrices.
They capture BFMs by the medium-access-control (MAC) frame-capturing tool, such as Wireshark \cite{wireshark}, because WiFi access points (APs) and stations (STAs) exchange BFMs without encryption \cite{ieee11ac}.
In other words, BFMs are stored in MAC-layer frames and exchanged among APs and STAs.
Because of this simplicity of availability of BFMs, BFMs would be the alternative features for CSI toward WiFi sensing. 

However, BFM-based sensing has not been well studied, and its applicability in sensing tasks is unclear, whereas CSI-based sensing is well-studied in the literature such as human localization and tracking~\cite{wu2016widir}, and heart rate estimation~\cite{zeng2019farsense}.
Furthermore, there is a performance gap between BFM- and CSI-based sensing, e.g. the gap in the respiration estimation error~\cite{kanda2022respiratory,zeng2019farsense}.

One promising approach to study the performance of BFM as an alternative of CSI for sensing tasks is to directly recomposite the CSI from the BFM and evaluate the prediction errors~\cite{hanaharaCCNC}.
The intuition is that if one can predict CSI from the BFM, the aforementioned various CSI-sensing task is also possible based on BFM by orderly performing the recomposition of CSI and sensing based on the recomposited CSI.
In~\cite{hanaharaCCNC}, they demonstrated the feasibility of the recompositing CSI amplitude from BFM by a simulation dataset.
However, there still existed a recomposition error between the ground-truth and recomposited CSI, and the input information that was additionally needed to enhance the prediction accuracy was not fully determined.
Answering this question leads to further understanding the BFM as an alternative feature for CSI towards wireless sensing tasks.

\begin{table}[t]
    \centering
    \caption{Summary of CSI recomposition Frameworks}
    \label{tbl:comparison_framework}
    \begin{tabular}{cccc}
        \toprule
        Framework & Input & Output & Model \\
        \midrule
        \cite{hanaharaCCNC} & BFM & CSI amplitude & SMI-BFM \\
        (For comparison) & Image & CSI amplitude & SMI-image \\
        Proposed & Image and BFM & CSI amplitude & MMI \\
        \bottomrule
    \end{tabular}
\end{table}

To this end, we develop a vision-aided CSI recomposition framework, using images alongside BFM.
This is motivated by our key hypothesis that spatial information on propagation environments in camera images can aid in the compensation of the insufficiency of BFM to recomposite CSI. 
This work is devoted to drawing first-hand insights into this hypothesis in the following two perspectives.

First, in view of the fact that the recomposition involves the two different modalities (i.e., BFMs and images), we design a multimodal input (MMI) deep learning model.
A brief summary of the proposed and comparison frameworks are listed in TABLE~\ref{tbl:comparison_framework}.
The MMI model is constructed by two encoders, which we term as the BFM encoder and image encoder, and CSI decoder.
The encoders extract information on the propagation environments from the BFMs and camera images, respectively, and the CSI decoder integrates them and recomposites the CSI amplitude.
Leveraging this architecture, we compare its performance with that of single-modal input (SMI) models and demonstrate that the accuracy enhancement by additionally using camera images is feasible.

Secondly, to demonstrate the feasibility of the proposed CSI recomposition, we conduct experiments using IEEE 802.11ac devices.
Despite the simplicity of the availability of BFMs, it was difficult to simultaneously obtain the corresponding CSI and BFM using off-the-shelf devices to create the datasets of the MMI learning.
To address this issue, we introduce an experimental system obtaining corresponding CSI and BFM, which was developed in \cite{itahara2021beamforming}.
In this system, BFM are emulated by computing singular value decomposition (SVD) instead of capturing frames, where CSI is obtained by the customized firmware~\cite{gringoli2019free} installed in ASUS RT-AC86U WLAN AP.

The main contributions of this paper are summarized as follows:
\begin{itemize}
    \item
    Toward firmware-agnostic WiFi sensing, i.e., sensing without using CSI, we propose a CSI recomposition framework using camera images in addition to BFMs.
    This framework is devoted to demonstrating that BFM may not involve sufficient spatial information on propagation environments for CSI recomposition, which could be compensated by camera images.
    To realize this framework, we design an MMI learning model with the input of the combination of BFM and image.
    \item
    We conduct experiments to demonstrate the feasibility of improving the CSI recomposition accuracy using an MMI model.
    The experimental results confirm that the recomposition accuracy of the proposed multimodal framework is more than that of the single-modal frameworks, using only images or BFMs.
\end{itemize}

The remainder of this paper is organized as follows.
The system model is introduced in Section~\ref{sec:system_model}.
Section~\ref{sec:method} presents the vision-aided CSI recomposition framework.
Section~\ref{sec:experimental_evaluation} presents the experimental evaluation.
Finally, Section~\ref{sec:conc} concludes the paper.

\section{System Model}
\label{sec:system_model}
\begin{figure}[t]
    \centering
    \includegraphics[width=.9\linewidth]{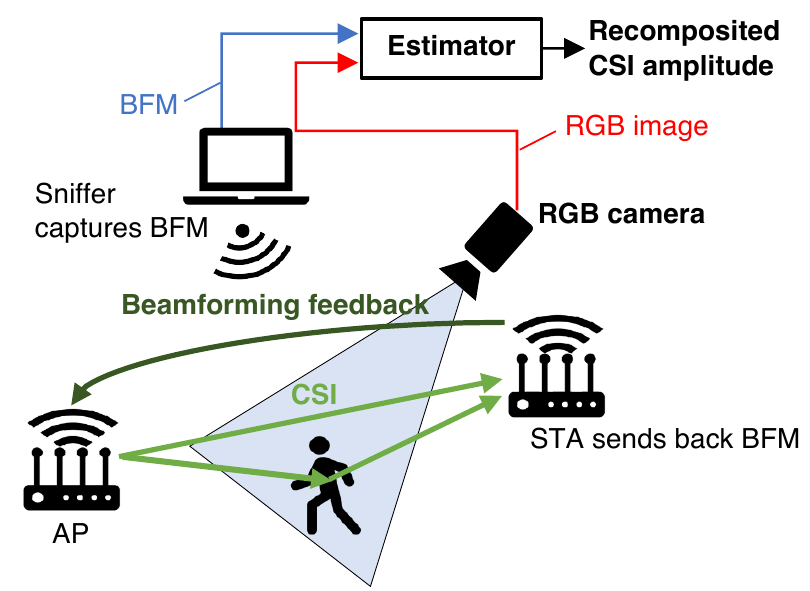}
    \caption{
        System model.
        Estimator recomposites CSI amplitude using the BFMs obtained from the sniffer and the RGB images obtained by the RGB camera.
    }
    \label{fig:system_model}
\end{figure}

Fig.~\ref{fig:system_model} presents a system model of our proposed framework, which consists of an AP with $M$ antennas, STA with $N$ antennas, red-green-blue (RGB) camera, sniffer, and estimator.
The STA receives frames transmitted by the AP, measures the CSI, calculates the BFM from the CSI, and sends back the BFM to the AP by transmitting beamforming feedback frames.
The sniffer captures the feedback frames to obtain the BFMs.
Because BFMs are sent without encryption, they can be captured by frame-capturing tools, such as Wireshark~\cite{wireshark}.

The estimator recomposites the amplitude of the CSI elements.
Unlike in a previous study~\cite{hanaharaCCNC}, the estimator uses not only BFMs but also RGB images captured by the RGB camera.
Let the CSI at time $t\in\mathbb{Z}$ and subcarrier $k\in\{1, 2, \dots, K\}$ be denoted by $\bm{H}_{t}[k]\in\mathbb{C}^{N\times M}$; then, the corresponding BFM $\bm{V}_{t}[k]\in\mathbb{C}^{M\times N}$ is given by:
\begin{align}
    \bm{H}_{t}[k] = \bm{U}_{t}[k]\,\bm{\varSigma}_{t}[k]\,\bm{V}_{t}[k]^{\mathrm{H}},
\end{align}
where $\bm{U}_{t}[k]$ and $\bm{V}_{t}[k]$ are unitary matrices, and $\bm{\varSigma}_{t}[k]$ is a diagonal matrix with singular values.
The notation $\bm{V}^{\mathrm{H}}$ denotes the Hermitian transpose of $\bm{V}$.
Let subcarrier-integrated CSI and BFM be denoted by $\{\bm{H}_{t}[k]\}_{k}$, $\{\bm{V}_{t}[k]\}_{k}$, respectively.
The RGB image denoted by $\bm{I}_{t}$ has the shape of $H\times W \times 3$, where $H$ and $W$ are the height and width of the image, respectively and 3 represents color channel, red, green and blue.
Using the data for $\{\bm{V}_{t}[k]\}_{k}$ and $\bm{I}_{t}$, the estimator recomposites the amplitude of the $\{\bm{H}_{t}[k]\}_{k}$.

RGB cameras are typically used as surveillance cameras in office rooms, where WLANs are employed.
Therefore, the cameras can be used for other applications such as surveillance to reduce the initial and running costs of deploying the cameras.

\section{Multimodal Learning-Based CSI Recomposition Framework}
\label{sec:method}
\begin{figure*}[t]
    \centering
    \includegraphics[width=0.72\linewidth]{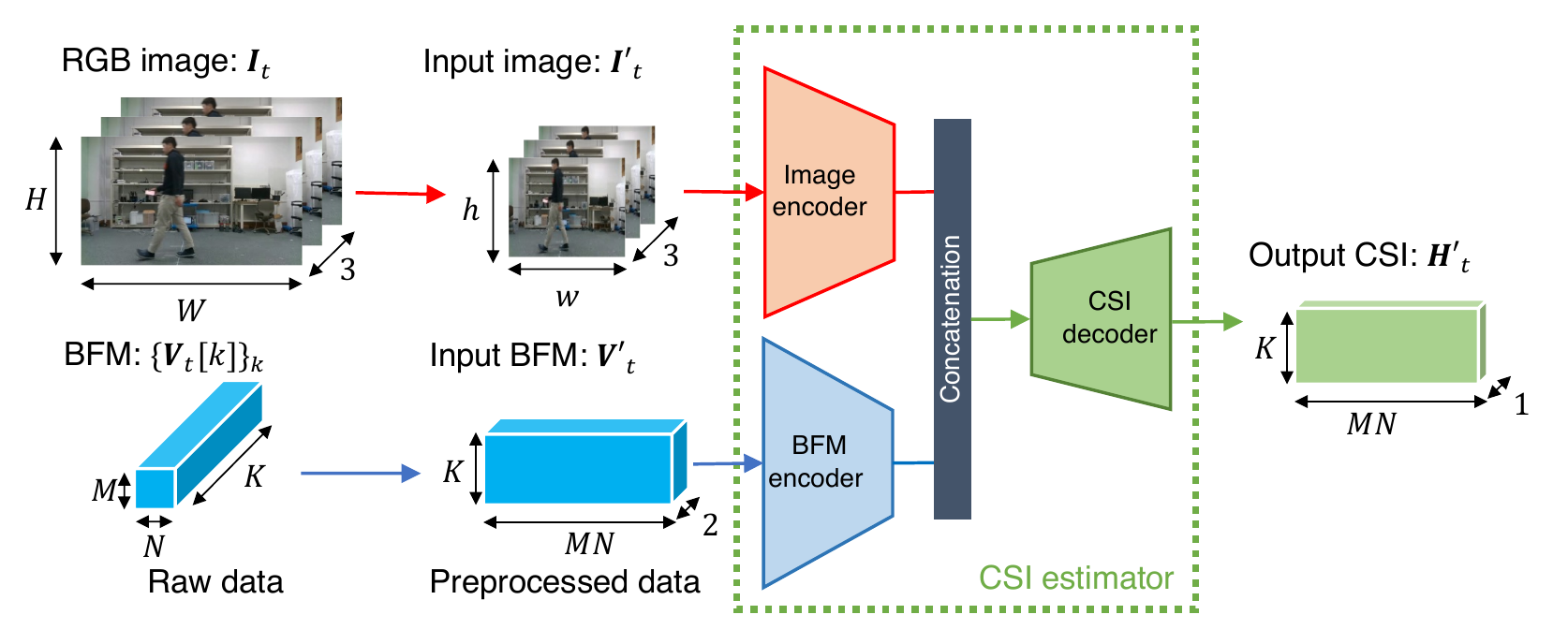}
    \caption{
        Overview of the proposed CSI recomposition framework.
        The CSI estimator is represented as the MMI model.
        Before being fed to the MMI model, first, the BFMs and RGB images are preprocessed.
        Multimodal inputs are concatenated at the concatenation layer to complement each other.
    }
    \label{fig:procedure_dataset}
\end{figure*}

\subsection{Data Preprocessing for CSI Recomposition}
\label{ssec:preprocessing}
The dataset for the CSI recomposition is generated through preprocessing, as shown in Fig.~\ref{fig:procedure_dataset}.
The dataset is composed of the input and supervised data.
The proposed CSI recomposition framework adopts an MMI model.
The BFMs are obtained by the frame-capturing tool, and the RGB images are obtained by the RGB camera.
Before being fed to the machine learning (ML) model, the obtained RGB image $\bm{I}_{t}$ is downsampled to $h \times w \times 3$, where $h \leq H$ and $w \leq W$, to reduce the calculation complexity.
Let the downsampled image be denoted by $\bm{I}'_{t}$.

The obtained BFM is preprocessed as follows:
First, the subcarrier-integrated BFM $\{\bm{V}_{t}[k]\}_{k}$ is flattened; namely, the shape of the BFM data is converted into $K \times MN \times 1$.
Subsequently, the absolute value and argument of the flattened complex BFM elements are calculated and concatenated to form the input BFM data $\bm{V}'_{t}$.
The shape of the input BFM data is $K \times MN \times 2$.

The supervised data, denoted by $\bm{H}'_{t}$, is the amplitude of the complex CSI elements corresponding to the input BFM elements.
The obtained CSI is preprocessed as follows:
First, as in the preprocessing of the BFMs, the obtained CSI is flattened to $K \times MN \times 1$.
Subsequently, each element of the flattened data is normalized.
The normalized data is used as the supervised data $\bm{H}'_{t}$ whose shape is $K \times MN \times 1$.

\subsection{Multimodal Input Model}
\label{ssec:learning_model}
This section details the MMI model used in the vision-aided CSI recomposition framework.
As previously mentioned, the preprocessed RGB image data $\bm{I}'_{t}$ and BFM data $\bm{V}'_{t}$ were fed to the MMI model.
The model was trained by computing the loss function between the output and supervised data $\bm{H}'_{t}$.

\begin{figure}[t]
    \centering
    \includegraphics[width=0.9\linewidth]{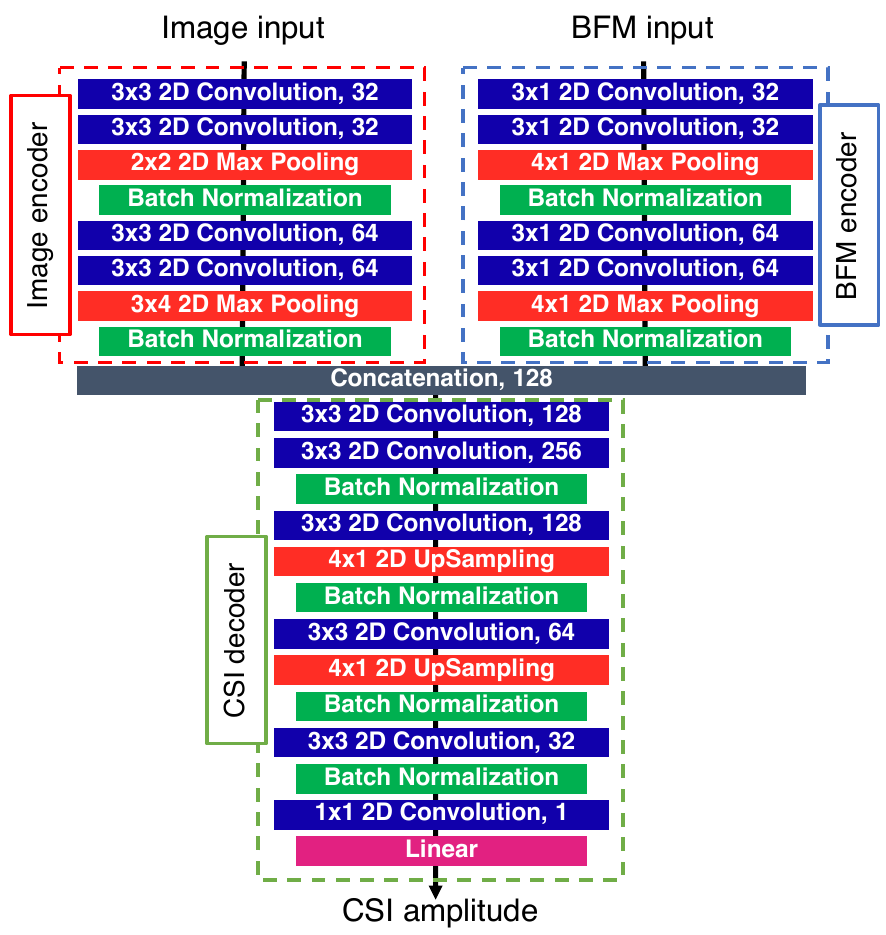}
    \caption{
        Structure of MMI model.
        The model consists of the BFM encoder and image encoder, CSI decoder, and concatenation layer.
    }
    \label{fig:learning_model}
\end{figure}

Fig.~\ref{fig:learning_model} shows the structure of the MMI model, which consists of two encoders, i.e., BFM encoder and image encoder, a concatenation layer, and CSI decoder.
The model is constructed as a convolutional neural network (CNN) including 0.92 million parameters.
This CNN consists of several layers:
2D convolution (2Dconv), 2D max pooling, batch normalization (BN)~\cite{ioffe2015batch}, concatenation, 2D upsampling, and linear layers.
Each 2Dconv layer is activated by the rectified linear unit except for the last 2Dconv layer.

The kernel and filter size of each layer are shown in Fig.~\ref{fig:learning_model}.
To ensure the consistency with the previous work~\cite{hanaharaCCNC}, the kernel size of the element domain is set to 1 in the BFM encoder, such that the 2Dconv and max pooling layers are utilized to extract inter-subcarrier features of each BFM.
The BFM and image encoder output features have the same shape.
In the concatenation layer, these features are concatenated and fed to the CSI decoder.
All 2Dconv layers conduct padding.
Finally, the CSI decoder is activated by a linear layer, and it outputs the recomposited CSI amplitude.

\section{Experimental Evaluation}
\label{sec:experimental_evaluation}

\subsection{Experimental Setup}
\label{ssec:setup_experiment}
\begin{figure}[t]
    \centering
    \subfloat[Equipment layout.]{
        \includegraphics[width=0.72\linewidth]{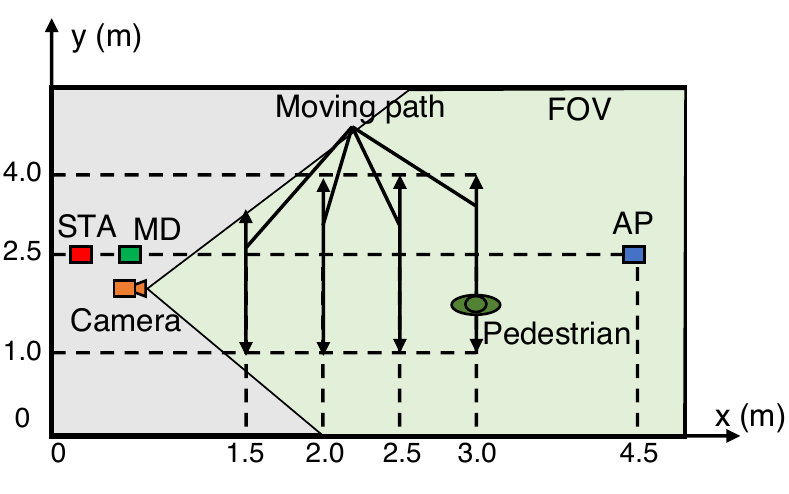}
        \label{subfig:experimental_environment_a}
    } \\
    \subfloat[Snapshot.]{
        \includegraphics[width=0.65\linewidth]{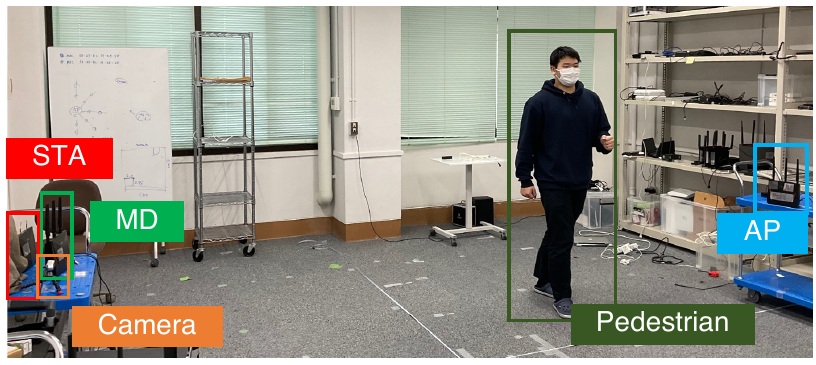}
        \label{subfig:experimental_environment_b}
    }
    \caption{
        Experimental system to obtain corresponding CSI and BFMs.
        The CSI captured by the MD represents the channel characteristic between the MD and STA.
        The MD behaves as the STA, which is just a receiver of the frames.
    }
    \label{fig:experimental_environment}
\end{figure}

This section details the experimental setups.
We develop an experimental system to obtain simultaneously corresponding CSI and BFM using off-the-shelf devices for creating the datasets.
Fig.~\ref{fig:experimental_environment} illustrates the experimental equipment layout, where an AP, STA, camera, and measurement device (MD) are used.
The MD captures frames transmitted by the AP to calculate the CSI.
Thus, the CSI represents the channel characteristic between the MD and AP.
In other words, the MD exhibited the characteristics of the STA, which was merely the receiver of the frames.
To obtain the corresponding CSI and BFM, the BFM of the input data is not obtained by the sniffer but obtained from the MD by applying SVD to the CSI.
Details of the experimental equipment are presented in TABLE~\ref{tbl:detail_equipment}.
The equipment is all off-the-shelf, except for the firmware of the MD, which is customized according to \cite{gringoli2019free}.
To experiment in a dynamic environment, a pedestrian as an obstacle moves along with the paths indicated in Fig.~\ref{fig:experimental_environment}\subref{subfig:experimental_environment_a}.
There are four paths perfectly included in the FOV of the camera.
We conduct a total of four experiments using each path.
The experimental parameters are presented in TABLE~\ref{tbl:experimental_parameters}.

The dataset with 24,000 samples of time-synchronized images and CSIs were obtained from the experimental data, and the corresponding BFMs were emulated from the CSIs.
The dataset is divided into three portions:
training, validation, and test data in the ratio of 72:18:10.
The ML model was trained using the training data.
The validation data was used to evaluate CNNs in each epoch.
The training is said to be complete when the number of training epochs reaches $100$
or when the validation loss increases by 10 epochs in a row,
which is referred to as early stopping and prevents the CNNs from overfitting.

\begin{table}[t]
    \caption{Details of Experimental Equipment}
    \label{tbl:detail_equipment}
    \centering
    \begin{tabular}{cc}
        \toprule
        Roles & Devices \\
        \midrule
        AP & ASUS RT-AC86U\\
        STA & ASUS RT-AC86U\\
        Measurement device & ASUS RT-AC86U with custom firmware\cite{gringoli2019free}\\
        RGB camera & Intel RealSense D435 \cite{IntelRealsense} \\
        \bottomrule
    \end{tabular}
\end{table}

\begin{table}[t]
    \caption{Experimental Parameters}
    \label{tbl:experimental_parameters}
    \centering
    \begin{tabular}{cc}
        \toprule
        Parameters & Values \\
        \midrule
        Number of transmit antennas $M$      & 3               \\
        Number of receive antennas $N$       & 4               \\
        Number of subcarrier index           & 256             \\
        Height of AP, STA, MD, and camera & $1.0\,\mathrm{m}$ \\
        \bottomrule
    \end{tabular}
\end{table}

\subsection{Baselines}

\begin{figure}[t]
    \centering
    \subfloat[SMI-BFM.]{
        \includegraphics[width=0.35\columnwidth]{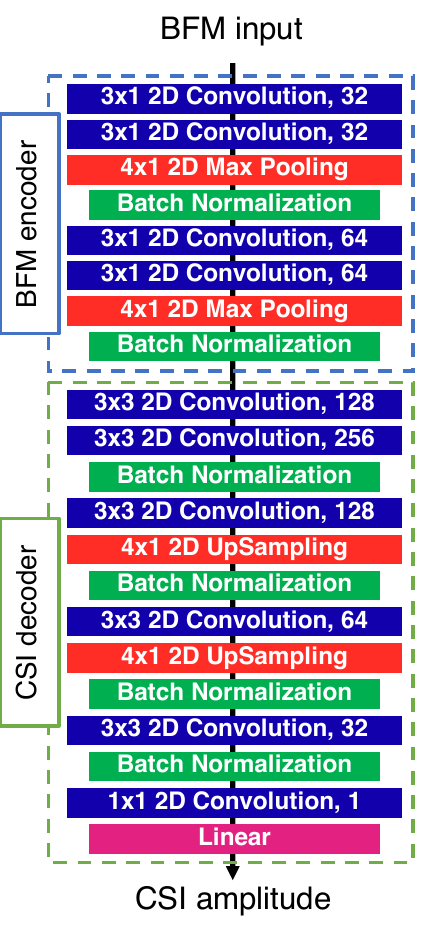}
        \label{subfig:learning_model_smibfm}
    }
    \subfloat[SMI-image.]{
        \includegraphics[width=0.35\columnwidth]{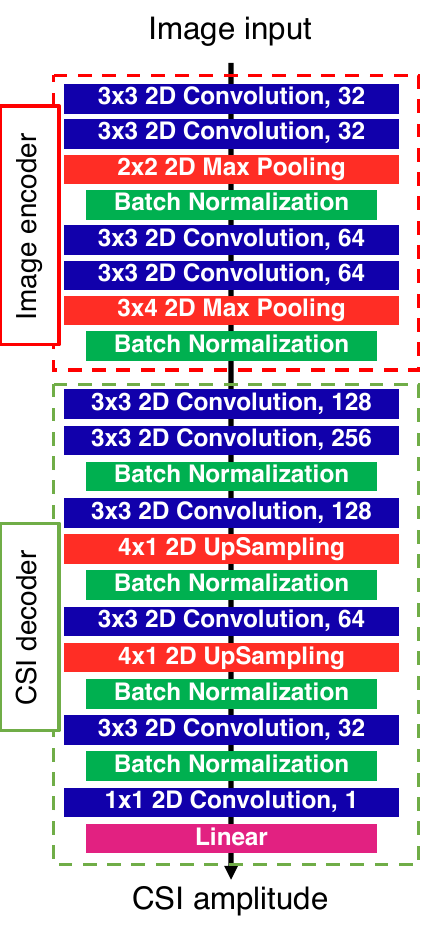}
        \label{subfig:learning_model_smiimage}
    }
    \caption{
        Structures of SMI-BFM and SMI-image models.
        SMI-BFM is constructed by the BFM encoder and CSI decoder, whereas the SMI-image is constructed by the image encoder and CSI decoder.
    }
    \label{fig:learning_model_baseline}
\end{figure}

To evaluate the performance of the MMI model, we used two SMI baselines:
SMI-BFM and SMI-image, whose inputs are the BFMs and images, respectively.
The structures of the SMI-BFM and SMI-image are shown in Fig.~\ref{fig:learning_model_baseline}.
They are constructed as CNN compliant with the MMI model, except for the concatenation layer and the layers before it.
They do not have the concatenation layer, and their first halves are the BFM and image encoder.
Their components are the same as the corresponding components in the MMI model.
The SMI-BFM and SMI-image model include 0.78 and 0.82 million parameters, respectively.

The hyperparameters of the ML algorithm are listed in TABLE~\ref{tbl:hyperparameters}.
We used Adam \cite{kingma2014adam} with a learning rate of 0.001 for the optimizer and mean squared error for the loss function.
To obtain the results without depending on the model initialization, all CNNs were trained based on five random seeds for initializing parameters.

\begin{figure*}[t]
    \centering
    \subfloat[$(1, 1)$ element.]{
        \includegraphics[width=.65\columnwidth]{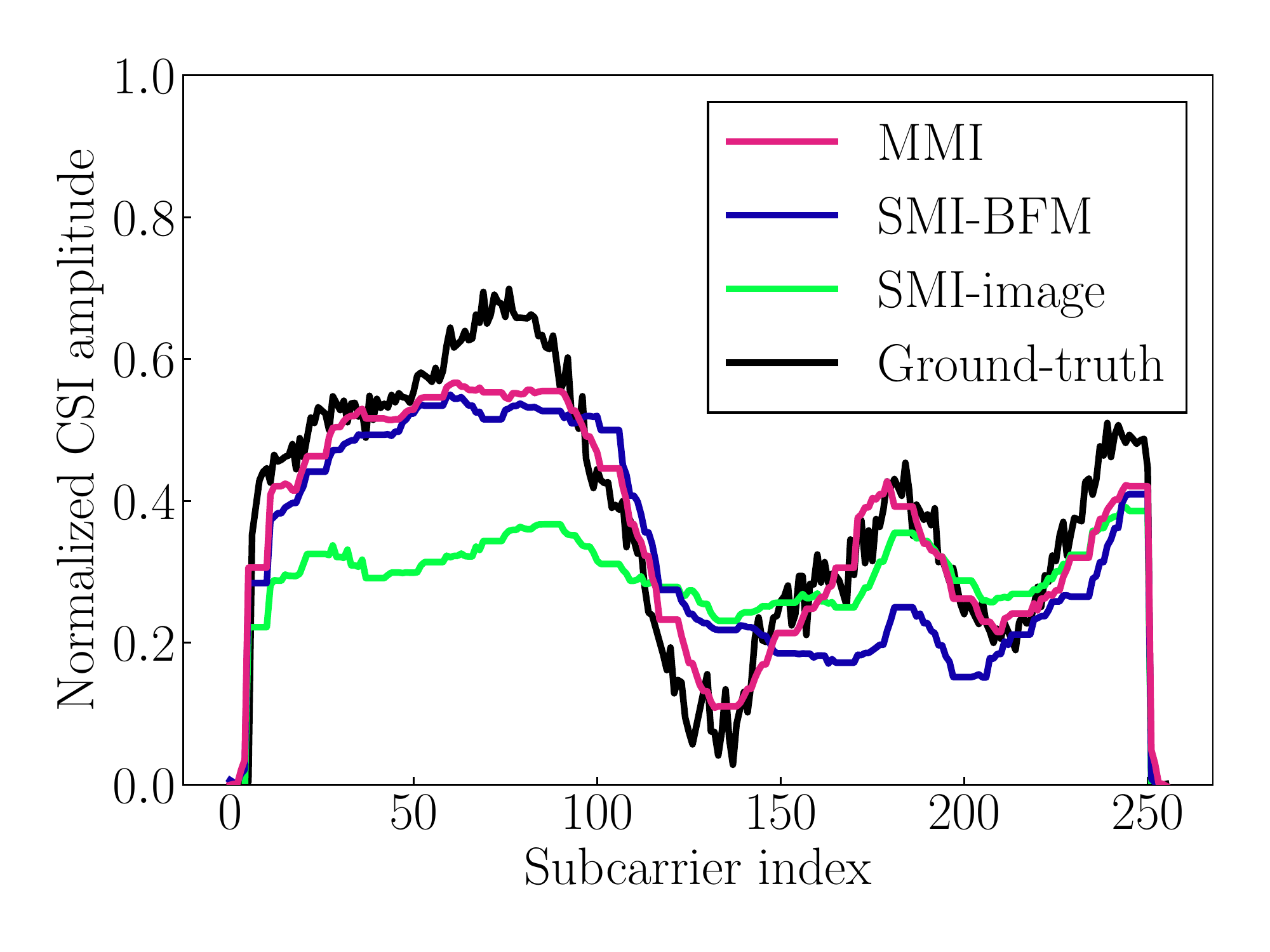}
        \label{subfig:csi_restore_1_1}
    }
    \subfloat[$(1, 2)$ element.]{
        \includegraphics[width=.65\columnwidth]{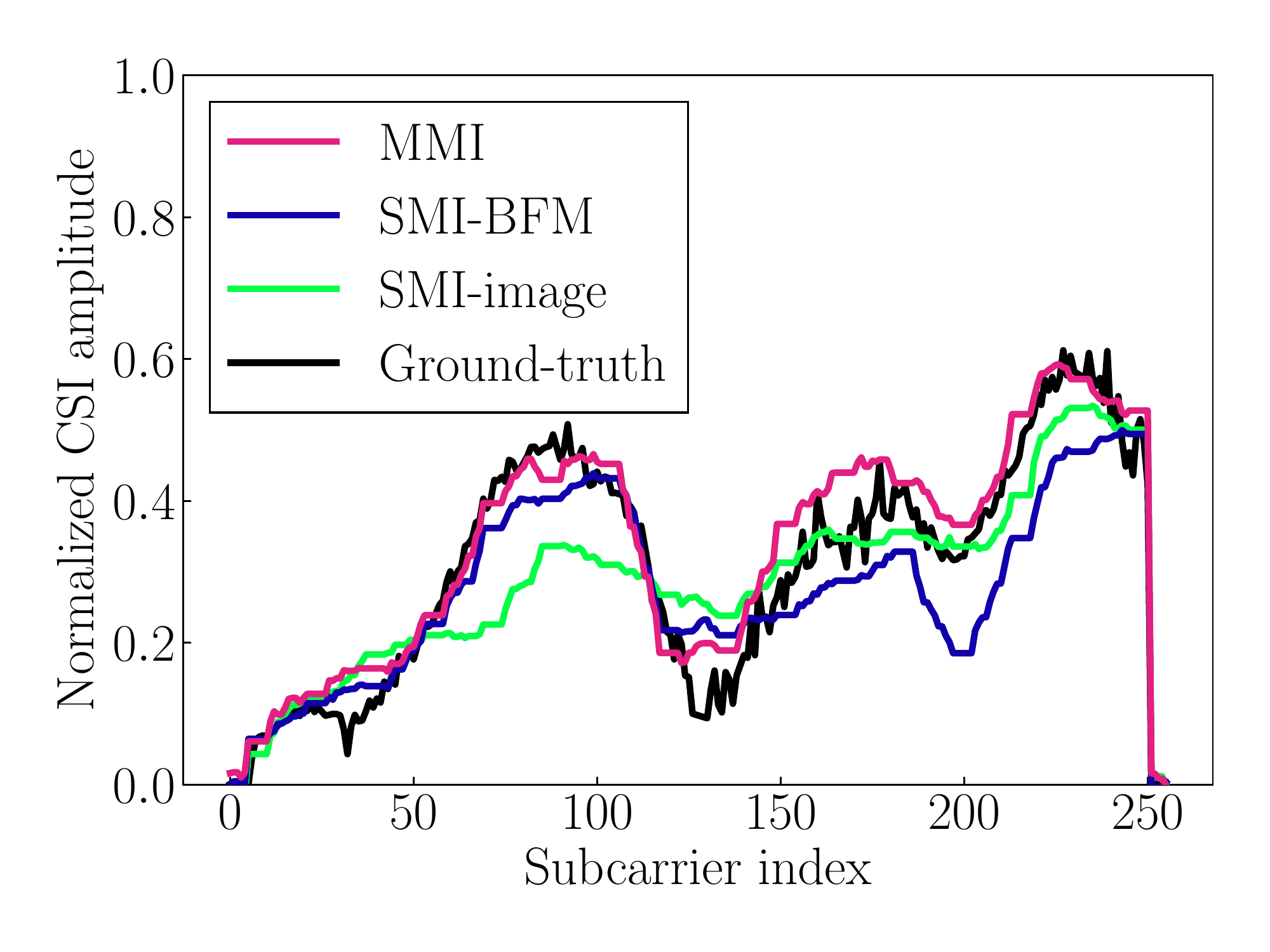}
        \label{subfig:csi_restore_2_1}
    }
    \subfloat[$(1, 3)$ element.]{
        \includegraphics[width=.65\columnwidth]{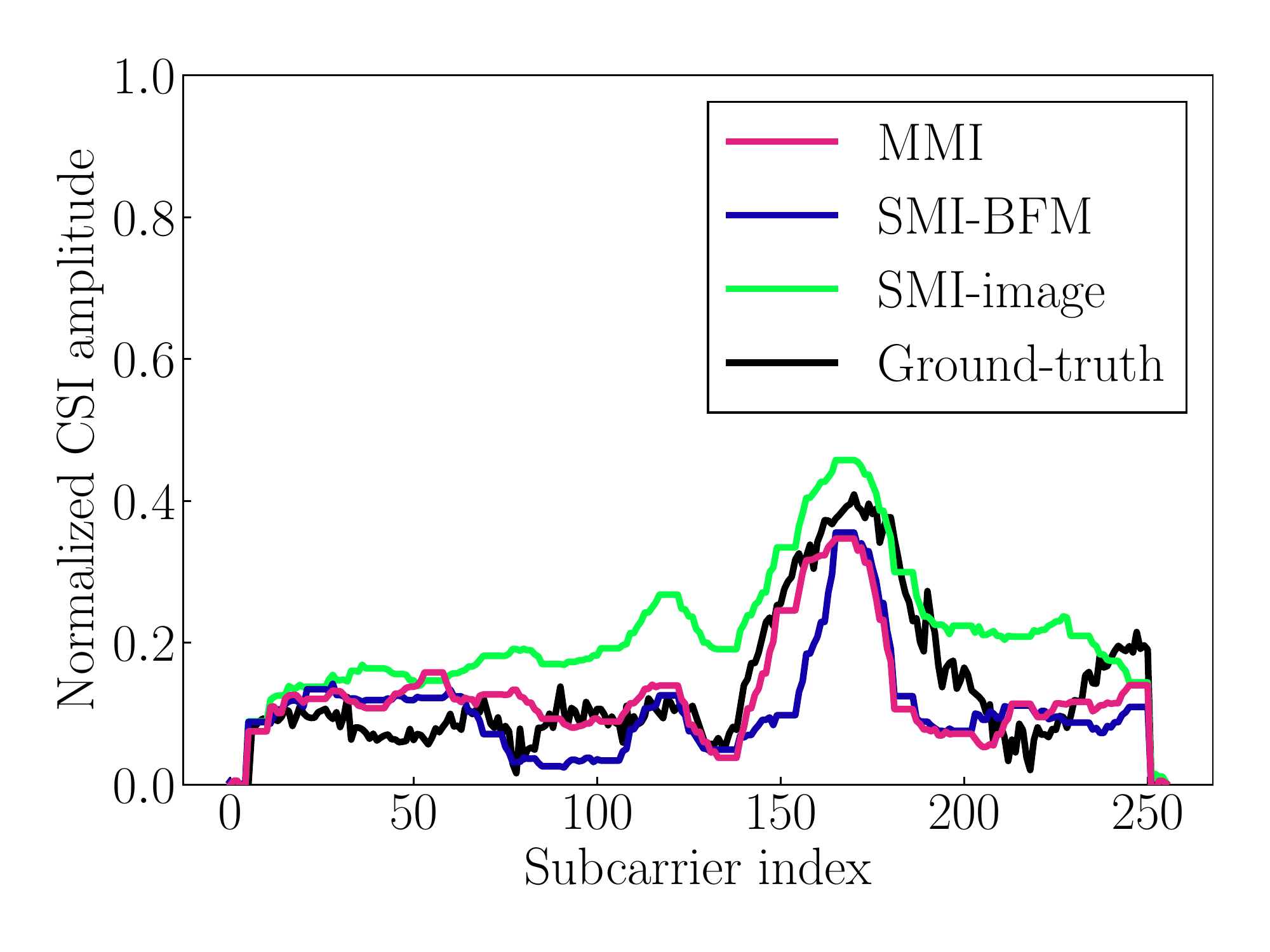}
        \label{subfig:csi_restore_3_1}
    }
    \caption{
        Example of the recomposited CSI and ground truth.
        The graph is depicted about the amplitude of the $(1, 1)$, $(1, 2)$, and $(1, 3)$ elements of $\bm{H}'_{t}$.
    }
    \label{fig:example_csi_recovery}
\end{figure*}

\begin{table}[t]
    \caption{Hyperparameters of Training}
    \label{tbl:hyperparameters}
    \centering
    \begin{tabular}{cc}
        \toprule
        Hyperparameters & Values \\
        \midrule
        Batch size & 64 \\
        Epochs & 100 \\
        Downsampled image pixels ($h \times w$) & $96 \times 96$ \\
        \bottomrule
    \end{tabular}
\end{table}

\subsection{Results}
\label{ssec:results}
Fig.~\ref{fig:example_csi_recovery} depicts an example of the frequency series of the amplitude of the CSI elements.
The recomposition accuracy is visually confirmed by this figure.
Fig.~\ref{fig:example_csi_recovery}\subref{subfig:csi_restore_1_1} shows that the recomposition of the MMI matches the ground truth the most compared with that of SMI-BFM.
However, from Fig.~\ref{fig:example_csi_recovery}\subref{subfig:csi_restore_2_1} and \subref{subfig:csi_restore_3_1}, there are some elements in which the accuracy gap between the MMI and SMI-BFM model is not high.

TABLE~\ref{tbl:rmse} shows the recomposition root-mean-squared error (RMSE) with one standard deviation for five random seeds.
The RMSE values in TABLE~\ref{tbl:rmse} are averaged over the samples.
From TABLE~\ref{tbl:rmse}, the recomposition error in the MMI is smaller than that in the baselines.
This result validates the feasibility of improving the recomposition accuracy using vision-aided CSI recomposition.
Particularly, compared with SMI-BFM, it is implied that the RGB images should include the spatial information on the propagation environments to fill the gap between the BFM and CSI.

\begin{table}
    \caption{Recomposition Errors}
    \label{tbl:rmse}
    \centering
    \begin{tabular}{lc}
        \toprule
        Model & RMSE \\
        \midrule
        SMI-image & 0.139 $\pm$ 0.000878 \\
        SMI-BFM & 0.108 $\pm$ 0.000627 \\
        \textbf{MMI} & \textbf{0.104} $\pm$ \textbf{0.000235} \\
        \bottomrule
    \end{tabular}
\end{table}

\section{Conclusions}
\label{sec:conc}
In this study, we proposed a CSI recomposition framework leveraging camera images alongside the BFMs.
The key idea was that the spatial information on the propagation environments included in the camera images could fill the gap between the BFM and CSI and improve the recomposition accuracy.
To leverage the camera images, we used MMI deep learning, where the input of the BFMs and camera images were integrated to recomposite the CSI amplitude.
We conducted experiments using IEEE 802.11ac devices to demonstrate the feasibility of the proposed CSI recomposition framework.
The experimental results revealed that the recomposition accuracy of the proposed multimodal framework was improved rather than that of the comparison single-modal frameworks.

\section*{Acknowledgment}
This research and development work was supported in part by the MIC/SCOPE \#JP196000002 and JSPS KAKENHI Grant Number JP18H01442.

\bibliographystyle{IEEEtran.bst}
\bibliography{globecom2022.bib}

\end{document}